\documentclass[a4paper,hyper,cits]{PoS-mod}
\usepackage{url,amsmath,amssymb,multirow}
\usepackage{subfigure}

\title{MoEDAL: Seeking magnetic monopoles and more at the LHC}

\ShortTitle{MoEDAL: Seeking magnetic monopoles and more at the LHC}

\author{\speaker{Vasiliki A.\ Mitsou} on behalf of the MoEDAL Collaboration\\
        Instituto de F\'isica Corpuscular (IFIC), CSIC -- Universitat de Val\`encia, \\ 
Parc Cient\'ific de la U.V., C/ Catedr\'atico Jos\'e Beltr\'an 2, \\
E-46980 Paterna (Valencia), Spain\\
and\\
CERN, PH Department, CH-1211 Geneva 23, Switzerland\\
        E-mail: \email{vasiliki.mitsou@ific.uv.es}}

\abstract{The MoEDAL experiment (Monopole and Exotics Detector at the LHC) is designed to directly search for magnetic monopoles and other highly ionising stable or metastable particles arising in various theoretical scenarios beyond the Standard Model. Its physics goals ---largely complementary to the multi-purpose LHC detectors ATLAS and CMS--- are accomplished by the deployment of plastic nuclear track detectors combined with trapping volumes for capturing charged highly ionising particles and TimePix pixel devices for monitoring. This paper focuses on the status of the detectors and the prospects for LHC Run~II.}

\FullConference{The European Physical Society Conference on High Energy Physics\\
		22--29 July 2015\\
		Vienna, Austria}

\begin{document}

\section{Introduction}\label{sc:intro}

MoEDAL (Monopole and Exotics Detector at the LHC)~\cite{moedal-web,moedal-tdr,jim}, the $7^{\rm th}$ experiment at the Large Hadron Collider (LHC)~\cite{LHC}, was approved by the CERN Research Board in 2010. It is designed to search for manifestations of new physics through highly-ionising particles in a manner complementary to ATLAS and CMS~\cite{DeRoeck:2011aa}. One of the primary motivations for the MoEDAL experiment is to pursue the quest for magnetic monopoles and dyons at LHC energies. Nonetheless the experiment is also designed to search for any massive, stable or long-lived, slow-moving particles~\cite{Fairbairn07} with single or multiple electric charges arising in many scenarios of physics beyond the Standard Model (SM)~\cite{Mitsou:2014dpa}. A selection of the physics goals and their relevance to the MoEDAL experiment are described here. For an extended and detailed account of the MoEDAL discovery potential, the reader is referred to the \emph{MoEDAL Physics Review}~\cite{Acharya:2014nyr}.

The structure of this paper is as follows. Section~\ref{sc:detector} provides a description of the different components of the MoEDAL detector with emphasis on the 2015 deployment. The physics program of MoEDAL is highlighted in Section~\ref{sc:physics}, focusing on magnetic monopoles and supersymmetric models predicting massive (meta)stable states. The paper concludes with an outlook in Section~\ref{sc:summary}.

\section{The MoEDAL detector}\label{sc:detector}

\begin{figure}[htb]
\begin{center}
\includegraphics[width=0.62\textwidth]{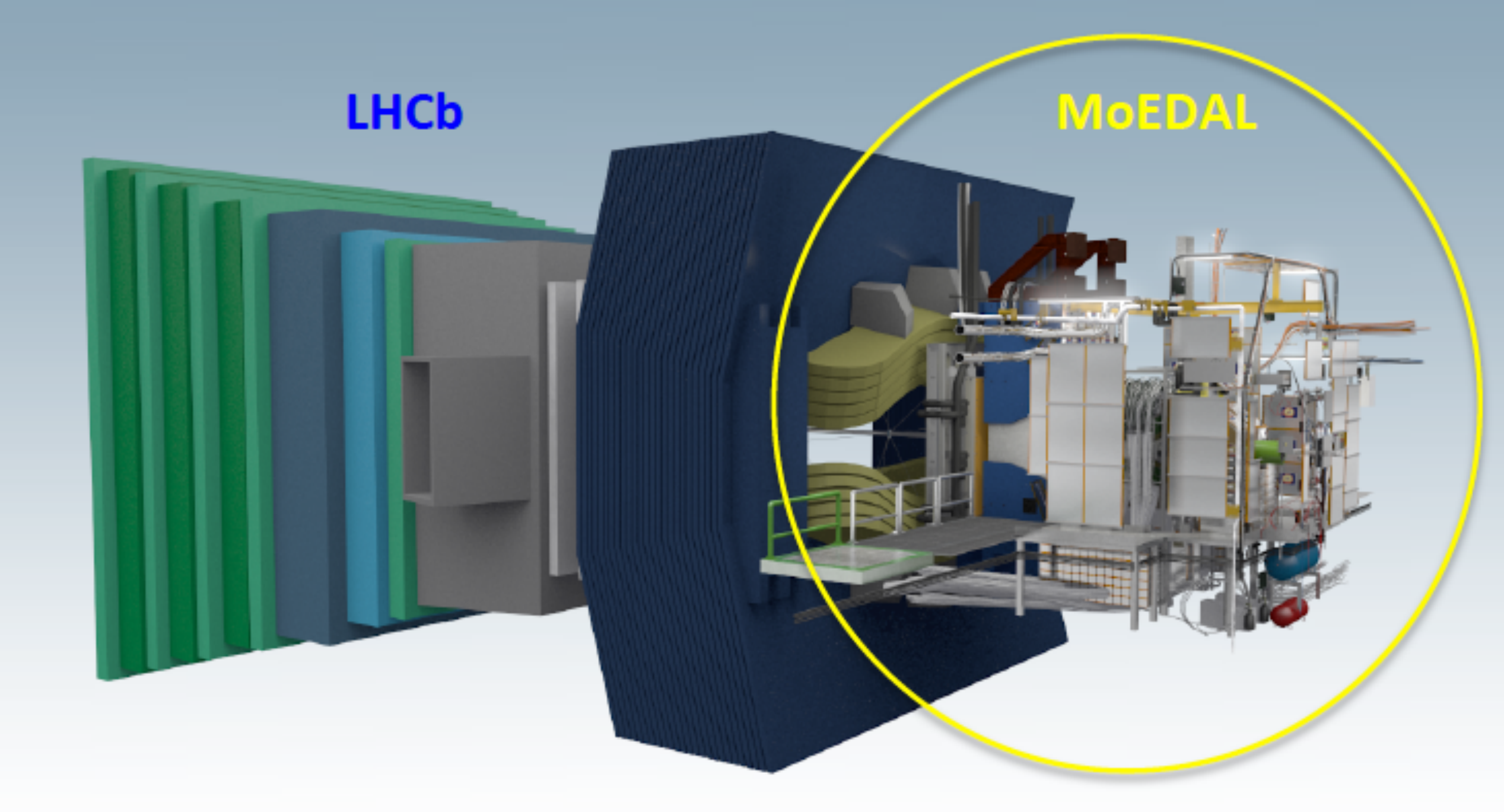}
\caption{ A three-dimensional schematic view of the MoEDAL detector (on the right) around the LHCb VELO region at Point~8 of the LHC.}
\label{fg:moedal}
\end{center}
\end{figure}

The MoEDAL detector~\cite{moedal-tdr} is deployed around the intersection region at Point~8 (IP8) of the LHC in the LHCb experiment Vertex Locator (VELO)~\cite{LHCb-detector} cavern. A three-dimensional depiction of the MoEDAL experiment is presented in Fig.~\ref{fg:moedal}. It is a unique and largely passive LHC detector comprised of four sub-detector systems. 

\subsection{Low-threshold nuclear track detectors}\label{sc:ndt}

\begin{figure}[htb]
\begin{minipage}[b]{0.55\textwidth}
\centering
\includegraphics[width=\textwidth]{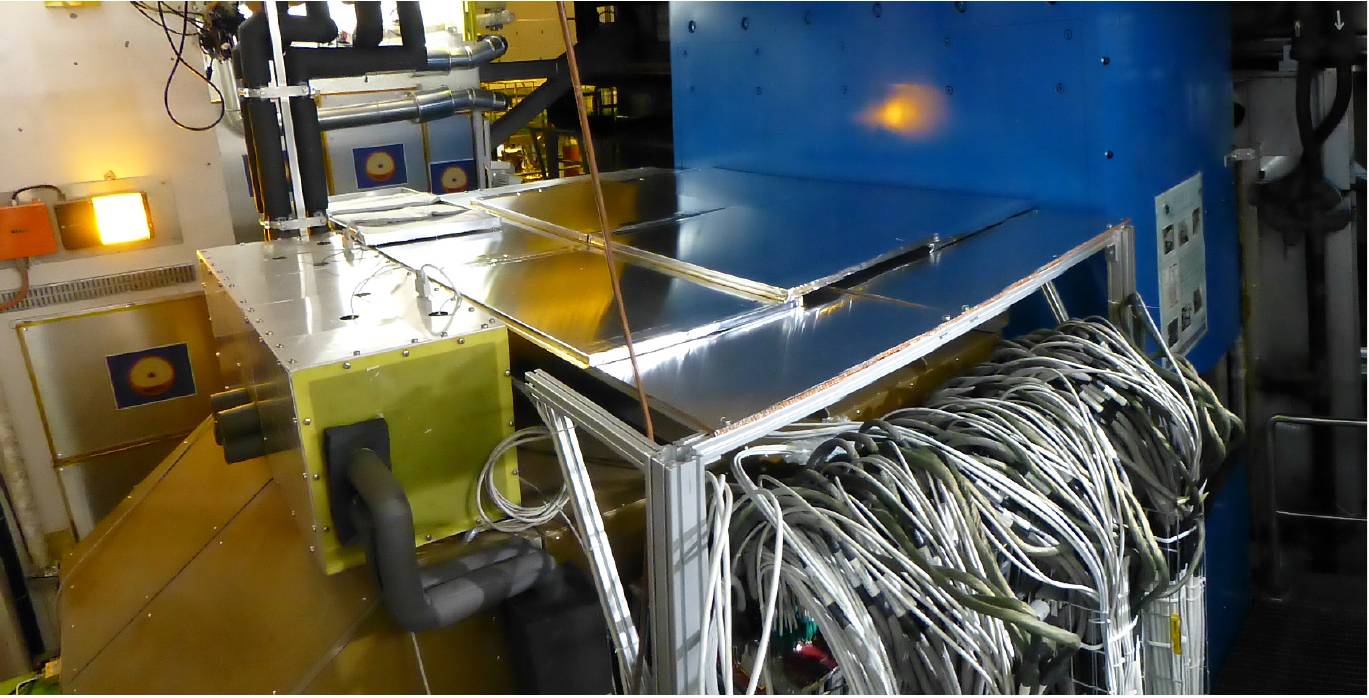}
\caption{\label{fg:ntd}Part of the 2015 NTD deployment on top of the LHCb VELO.}
\end{minipage}
\hfill
\begin{minipage}[b]{0.4\textwidth}
\centering
\includegraphics[width=0.8\textwidth]{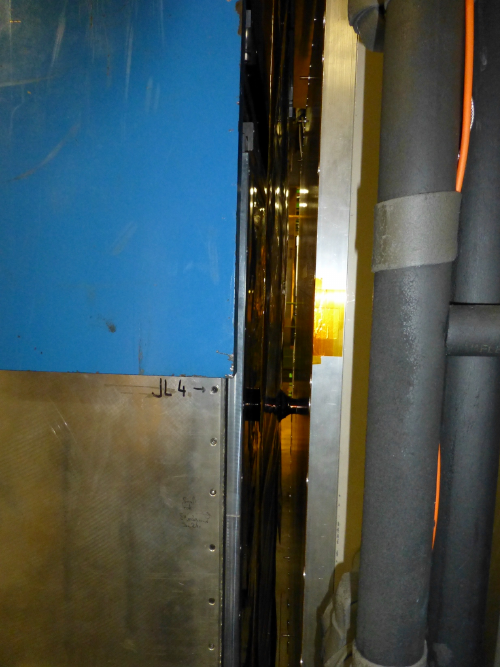}
\caption{\label{fg:vhcc}The VHCC between RICH1 and TT installed for the 2015 run.}
\end{minipage} 
\end{figure}

The main sub-detector system is made of a large array of CR39\textregistered,  Makrofol\textregistered\ and Lexan\textregistered\ nuclear track detector (NTD) stacks surrounding the intersection area. The passage of a highly-ionising particle through the plastic detector is marked by an invisible damage zone along the trajectory. The damage zone is revealed as a cone-shaped etch-pit when the plastic detector is etched using a hot sodium hydroxide solution. Then the sheets of plastics are scanned looking for aligned etch pits in multiple sheets. The MoEDAL NTDs have a threshold of $Z/\beta\sim5$, where $Z$ is the charge and $\beta=v/c$ the velocity of the incident particle. In proton-proton collision running, the only source of known particles that are highly ionising enough to leave a track in MoEDAL NTDs are spallation products with range that is typically much less than the thickness of one sheet of the NTD stack. In that case the ionising signature will be that of a very low-energy electrically-charged \emph{stopped} particle. This signature is distinct to that of a \emph{penetrating} electrically or magnetically charged particle that will usually traverse every sheet in a MoEDAL NTD stack, accurately demarcating a track that points back to the collision point with a resolution of $\sim 1~{\rm cm}$. The part of the 2015 NTD deployment which rests on top of the LHCb VELO is visible in Fig.~\ref{fg:ntd}. This is the closest possible location to the interaction point and represents a novelty of the 2015 run with respect to earlier installations. 

\subsection{Very high-charge catcher}\label{sc:vhcc}

Another novel feature of the 2015 deployment is the installation of a high-threshold NTD array ($Z/\beta\sim50$): the Very High Charge Catcher (VHCC).  The VHCC subdetector, consisting of two flexible low-mass stacks of Makrofol\textregistered\ in an aluminium foil envelope, is deployed in the forward acceptance of the LHCb experiment between the LHCb RICH1 detector and the Trigger Tracker (TT), as shown in Fig.~\ref{fg:vhcc}. It is the only NTD (partly) covering the forward region, adding only $\sim0.5\%$ to the LHCb material budget while enhancing considerably the overall geometrical coverage of MoEDAL NTDs.

\subsection{Magnetic trappers}\label{sc:mmt}

\begin{figure}[htb]
\begin{minipage}{0.475\textwidth}
\centering
\includegraphics[width=0.85\textwidth]{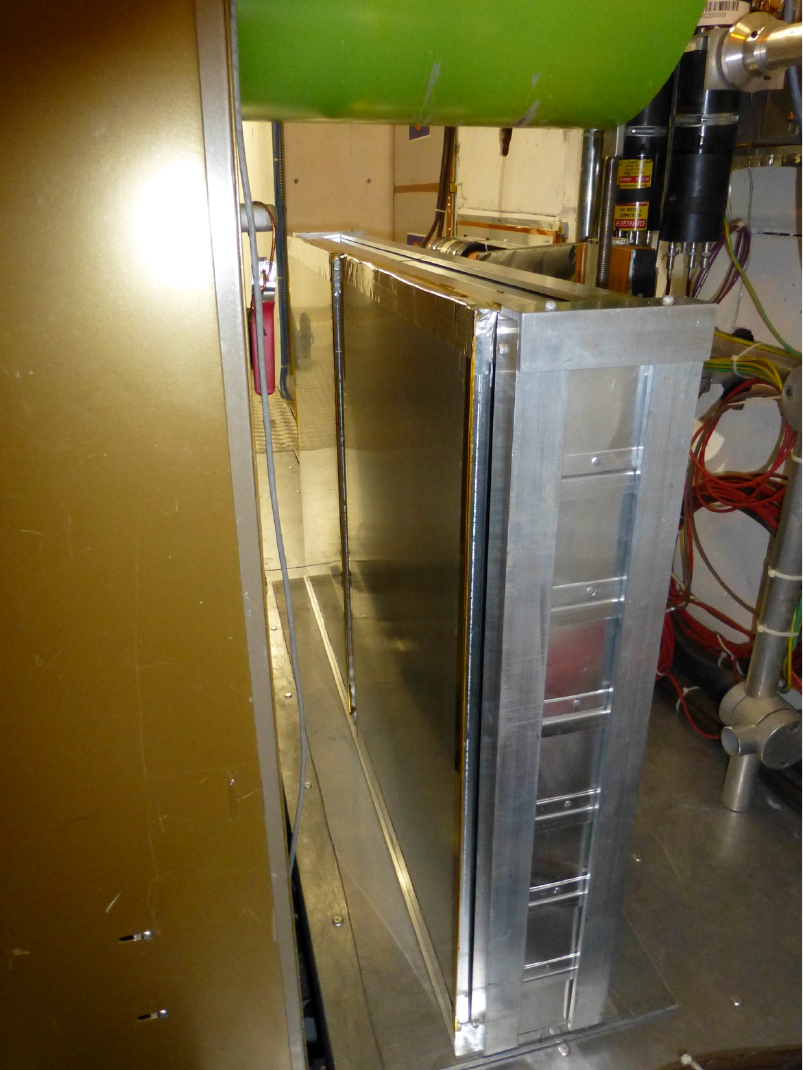}
\caption{\label{fg:mmt}Deployment of the MMT for the 2015 LHC run.}
\end{minipage}
\hfill
\begin{minipage}{0.475\textwidth}
\centering
\includegraphics[width=0.85\textwidth]{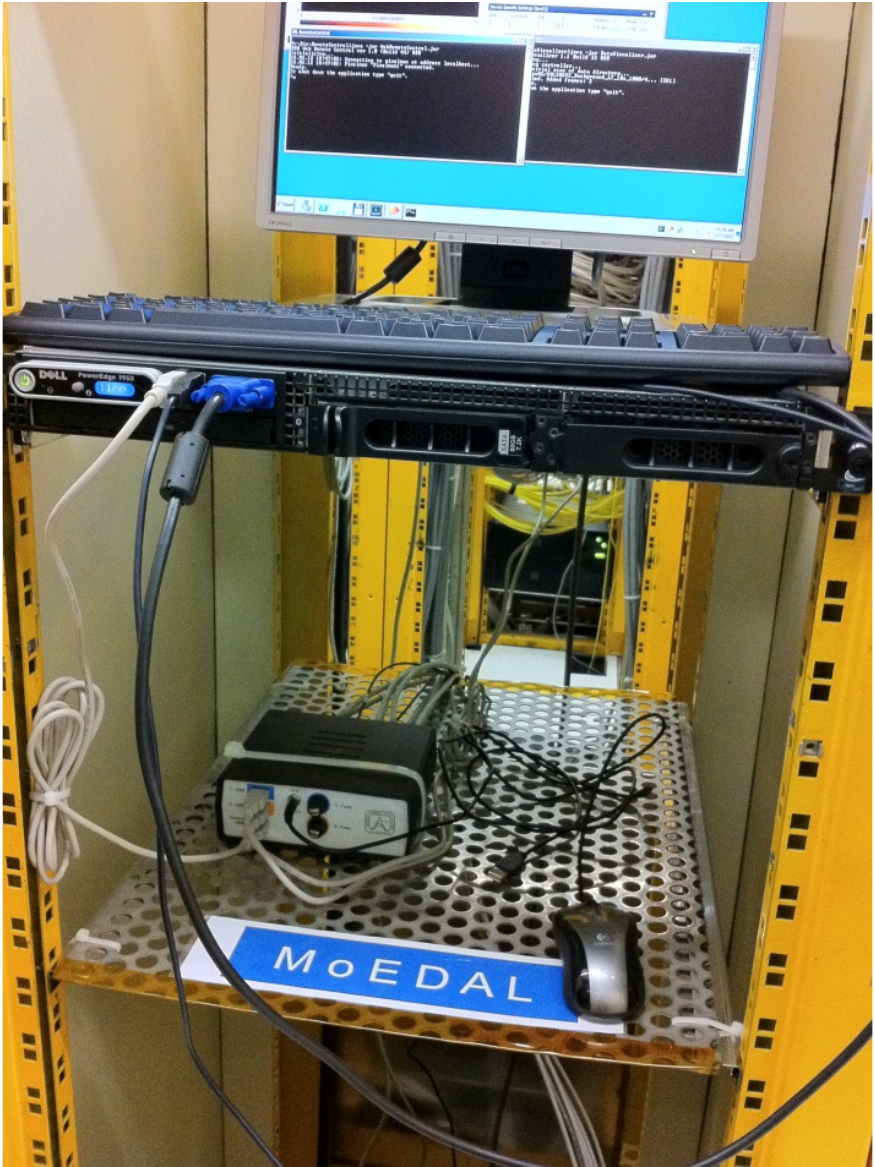}
\caption{\label{fg:timepix}2015 deployment of TimePix chips in MoEDAL.}
\end{minipage} 
\end{figure}

A unique feature of the MoEDAL detector is the use of paramagnetic magnetic monopole trappers (MMTs) to capture electrically- and magnetically-charged highly-ionising particles. Such volumes installed in IP8 for the 2015 proton-proton collisions is shown in Fig.~\ref{fg:mmt}. The aluminium absorbers of MMTs are subject to an analysis looking for magnetically-charged particles at a remote SQUID magnetometer facility~\cite{Joergensen:2012gy,DeRoeck:2012wua}. The search for the decays of long-lived electrically charged particles that are stopped in the trapping detectors will subsequently be carried out at a remote underground facility. A trapping detector prototype has alreday been exposed to 8~TeV proton-proton collisions for an integrated luminosity of 0.75~fb$^{-1}$ in 2012. It comprised an aluminium volume consisting of 11~boxes each containing 18~cylindrical rods of 60~cm length and 2.5~cm diameter. The first results interpreted in terms of monopole mass and magnetic charge from this run are going to be made public very soon~\cite{mmt2012}.

\subsection{Timepix radiation monitors}\label{sc:timepix}

The only non-passive MoEDAL sub-detector system comprises an array of TimePix pixel device arrays ($256\times256$ square pixels with a pitch of $55~{\rm \mu m}$)  distributed  throughout  the  MoEDAL cavern at IP8, forming a real-time radiation monitoring system of highly-ionising beam-related  backgrounds. A photo of its readout setup for the 2015 installations is shown in Fig.~\ref{fg:timepix}. Each pixel of the innovative TimePix chip comprises a preamplifier, a discriminator with threshold adjustment, synchronization logic and a 14-bit counter. The operation of TimePix in time-over-threshold mode allows a 3D mapping of the charge spreading effect in the whole volume of the silicon sensor, thus differentiating between different types of particles species from mixed radiation fields and measuring their energy deposition~\cite{timepix}.

\section{Physics goals}\label{sc:physics}

The search for highly-ionising long-lived particles with MoEDAL focuses on massive magnetically charged particles such as the magnetic monopole or the dyon. Besides that any other state that lives long enough to reach the NTDs and is slow moving ($\beta\lesssim 0.5$) and/or has multiple electric charge is of interest to MoEDAL. The general-purpose LHC detectors, ATLAS and CMS, have also performed searches for such particles however they are not optimised for their signatures~\cite{atlas-monopole}. They typically require a rather large statistical sample to establish a signal in the presence of usually considerable SM backgrounds. MoEDAL, on the other hand, has no trigger or electronic readout, thus, slowly moving particles do not pose any problem and only a few candidate events are enough to confirm a signal with no SM backgrounds to fake a beyond-SM signal. 

\subsection{Magnetic monopoles}\label{sc:mm}

The MoEDAL detector is designed to fully exploit the energy-loss mechanisms of magnetically charged particles~\cite{Dirac1931kp,Diracs_idea,tHooft-Polyakov,Cho1996qd}  in order to optimise its potential to discover these messengers of new physics. There are various theoretical scenarios in which magnetic charge would be produced  at the LHC~\cite{Acharya:2014nyr}: (light) 't Hooft-Polyakov monopoles~\cite{tHooft-Polyakov,Vento2013jua,arttu}, electroweak monopoles~\cite{Cho1996qd} and monopolium~\cite{Diracs_idea,khlopov,Monopolium,Monopolium1}. Magnetic monopoles that carry a non-zero magnetic charge and dyons possessing both magnetic and electric charge are among the most fascinating  hypothetical particles. Even though there is no generally acknowledged empirical  evidence for their existence, there are strong theoretical reasons to believe that they do exist, and they are predicted by many theories including grand unified theories and superstring theory~\cite{Rajantie:2012xh}. A recent review on existing experimental limits and future prospects for monopole discovery can be found in Ref.~\cite{laura}.

No magnetically-charged particles have been observed so far. A possible explanation for this lack of experimental confirmation is Dirac's proposal~\cite{Dirac1931kp,Diracs_idea,khlopov} that monopoles are not seen freely because they form a bound state called \emph{monopolium}~\cite{Monopolium,Monopolium1,Epele0} being confined by strong magnetic forces. Monopolium is a neutral state, hence it is difficult to detect directly at a collider detector, although its decay into two photons would give a rather clear signal for the ATLAS and CMS detectors~\cite{Epele1}, which however would not be visible in the MoEDAL detector. Nevertheless the monopolium might break up in the medium of MoEDAL into highly-ionising dyons, which subsequently can be detected in MoEDAL~\cite{Acharya:2014nyr}. Moreover its decay via photon emission would produce a peculiar trajectory in the medium, if the decaying states are also magnetic multipoles~\cite{Acharya:2014nyr}.

\paragraph{D-matter}\label{sc:dm}

Some versions of string theory include higher-dimensional ``domain-wall''-like membrane \emph{(D-brane)} structures in space-time. Fundamental open strings that represent particle excitations of the vacuum have their ends attached to these  D-branes. In some versions of string theory three-dimensional domain-wall D-branes are compactified and embedded in higher-spatial dimensional bulk spaces. If our world is viewed as a large membrane, then from a low-energy observer's perspective, such compactified three-dimensional structures would effectively appear to be point-like \emph{D-particles}, treated as quantum excitations above the vacuum~\cite{westmuckett,shiu} and collectively referred to as {\it D-matter}. D-matter states could be light enough to be phenomenologically relevant at the LHC.

D-particles, like all other D-branes, are solitonic non-perturbative objects in the string/brane theory with similarities and differences with magnetic monopoles with non-trivial cosmological implications~\cite{Witten2002wb,westmuckett,Mavromatos:2010jt}. 
An important difference between the D-matter states and other non-perturbative objects, such as magnetic monopoles, is that they could have {\it perturbative} couplings, with no magnetic charge in general. Nonetheless, in the context of brane-inspired gauge theories, brane states with magnetic charges can be constructed, which would manifest themselves in MoEDAL in a manner similar to  magnetic monopoles. 

\subsection{Supersymmetric long-lived particles}\label{sc:susy}

Supersymmetry (SUSY) is an extension of the Standard Model which provides elegant solutions to several open issues in the SM, such as the hierarchy problem, the identity of dark matter~\cite{mitsou-dm}, and the grand unification. No highly-charged particles are expected in this theoretical framework, however SUSY scenarios propose a number of massive slowly moving ($\beta \lesssim 0.5$)  electrically charged particles. If they are sufficiently long-lived to travel a distance of at least ${\cal O}(1~{\rm m})$  before decaying and their $Z/\beta\gtrsim 5$,  then they will be detected in the MoEDAL NTDs.

\paragraph{Metastable lepton NLSP in the CMSSM with a neutralino LSP}

The next-to-the-lightest supersymmetric particle (NLSP) may be one instance of long-lived sparticle. This would occur, for example, if the lightest supersymmetric particle (LSP) is the gravitino, or if the mass difference between the NLSP and the neutralino LSP is small, offering more scenarios for long-lived charged sparticles. In {\it neutralino dark matter} scenarios based on the constrained MSSM (CMSSM), for instance, the most natural candidate for the NLSP is the lighter tau slepton ${\tilde \tau_1}$~\cite{stauNLSP}, which could be long lived if $m_{\tilde \tau_1} - m_{\tilde{\chi}_1^0}$ is small. There are several regions of the CMSSM parameter space compatible with the discovered Higgs boson mass and the relic density of the LSP~\cite{MC8}, while assuring a ${\tilde \tau_1}$ lifetime $\gtrsim 100$~ns, that it is likely to escape the detector before decaying, and hence would be detectable as a massive, slowly-moving charged particle~\cite{Sato}. 

\paragraph{Metastable stops in gravitino LSP scenarios}

\begin{figure}[htb]
\begin{minipage}[b]{0.475\textwidth}
\centering
\includegraphics[width=\textwidth]{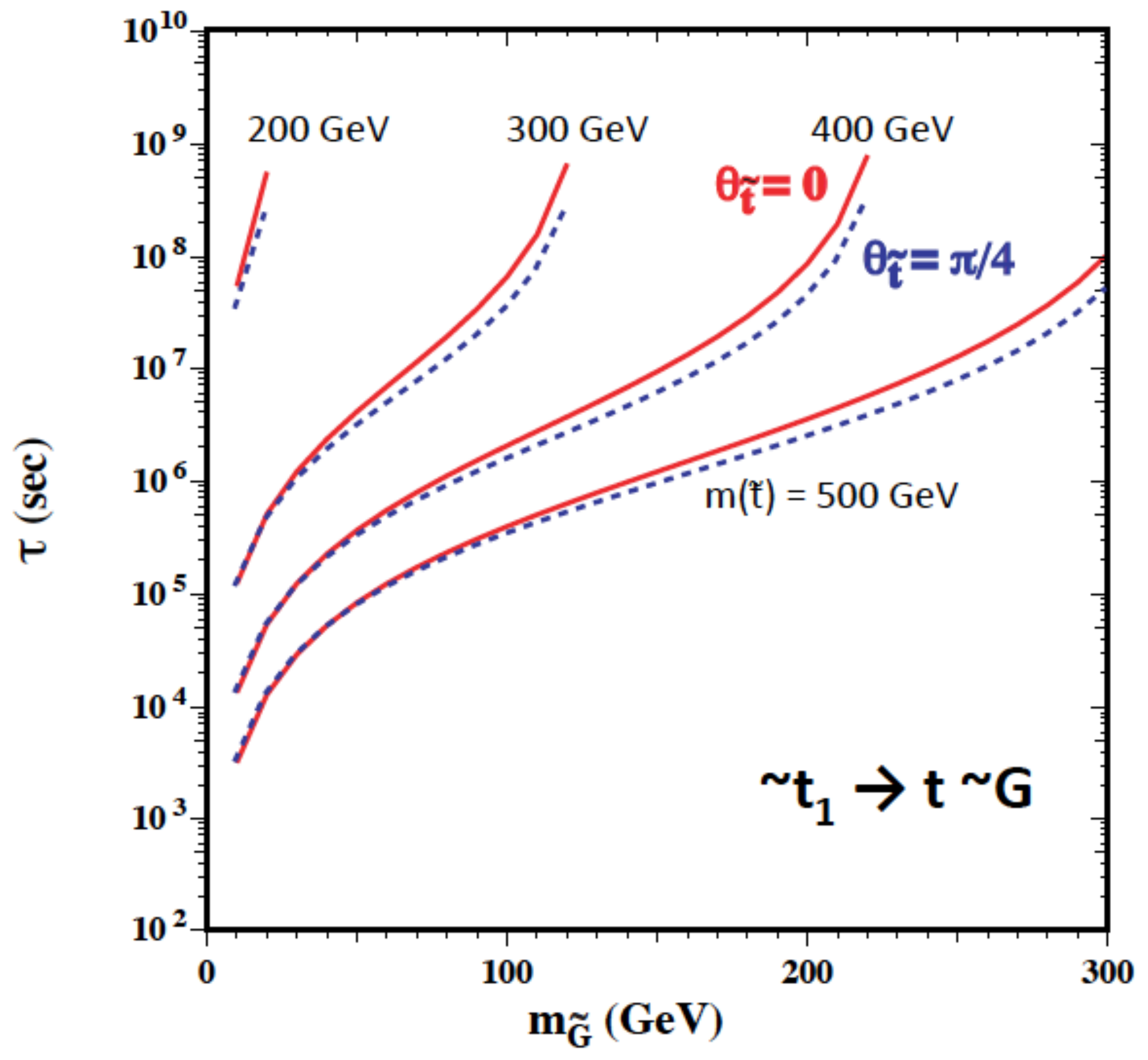}
\end{minipage}
\hfill
\begin{minipage}[b]{0.475\textwidth}
\centering
\caption{The stop lifetime as a function of $m_{\tilde G}$ for $m_{\tilde{t}} = 200, 300, 400$ and $500~{\rm GeV}$, shown for the case of the two-body decay ${\tilde t_1} \to {\tilde G} t$,
the dominant mode for $m_{\tilde G} < m_{\tilde{t}} - m_t$, assuming zero stop mixing (red solid line) and maximal mixing (blue dashed line). Adapted from Ref.~\cite{stopNLSP}.}
\label{fig:susy3}
\end{minipage} 
\end{figure}

In {\it gravitino dark matter} scenarios with more general options for the pattern of supersymmetry breaking, other options appear quite naturally, including the lighter tau slepton ${\tilde \tau_1}$~\cite{CAPTURE3}, or a sneutrino${\tilde \nu}$~\cite{sleptonNLSP}, or the lighter top squark ${\tilde t_1}$~\cite{stopNLSP}. If the gravitino ${\tilde G}$ is the LSP, as we can see in Fig.~\ref{fig:susy3}, the stop lifetime is relatively insensitive to the stop mixing angle $\theta_{\tilde{t}}$, yet depends on the sparticle masses, and spans the range $10^3 - 10^9~{\rm s}$. Clearly, this is extremely long compared with the QCD hadronisation time-scale, so that the stop NLSP forms metastable colour-singlet states, the so-called \emph{R-hadrons}, detectable in the MoEDAL detector.

\paragraph{Long-lived gluinos in split supersymmetry}

The above discussion has been in the context of scenarios where all the supersymmetric partners of Standard Model particles have masses in the TeV range. Another scenario is ``split supersymmetry'', in which the supersymmetric partners of quarks and leptons are very heavy, of a scale $m_s$, whilst the supersymmetric partners of SM bosons are relatively light~\cite{splitSUSY}. In such a case, the gluino could have a mass in the TeV range and hence be accessible to the LHC, but would have a very long lifetime:
\begin{equation}
\tau \approx 8 \left( \dfrac{m_s}{10^9~{\rm GeV}} \right)^4 \left( \dfrac{1~{\rm TeV}}{m_{\tilde{g}}} \right)^5~{\rm s}.
\label{gluinotau}
\end{equation}
Long-lived gluinos would form gluino R-hadrons including gluino-gluon (gluinoball) combinations, gluino-$q{\bar q}$ (mesino) combinations and gluino-$qqq$ (baryino) combinations. The heavier gluino hadrons would be expected to decay into the lightest species, which would be metastable, with a lifetime given by Eq.~(\ref{gluinotau}), and it is possible that this metastable gluino hadron could be charged. In the same way as stop hadrons, gluino hadrons may flip charge through conventional strong interactions as they pass through matter, and it is possible that one may pass through most of a conventional LHC tracking detector undetected in a neutral state before converting into a metastable charged state that could be observed with MoEDAL. 

\section{Conclusions and outlook}\label{sc:summary}

MoEDAL is going to extend considerably the LHC reach in the search for (meta)stable highly ionising particles. The latter are predicted in a variety of theoretical models, such as magnetic monopoles and SUSY long-lived spartners~\cite{Acharya:2014nyr}. The MoEDAL design is optimised to probe precisely such states, unlike the other LHC experiments~\cite{DeRoeck:2011aa}. Furthermore it combines different detector technologies: plastic nuclear track detectors, trapping volumes and pixel sensors~\cite{moedal-tdr}. The first physics results, obtained with the trapping volumes exposed to LHC Run~I collisions, are going to be published very soon~\cite{mmt2012}. The MoEDAL experiment is currently deployed and exposed to 13~TeV proton-proton collisions at the LHC looking for the least explored signals of New Physics.

\section*{Acknowledgements}

The author acknowledges support by the Spanish Ministry of Economy and Competitiveness (MINECO) under the project FPA2012-39055-C02-01, by the Generalitat Valenciana through the project PROMETEO~II/2013-017, by the Centro de Excelencia Severo Ochoa SEV-2014-0398 and by the Spanish National Research Council (CSIC) under the JAE-Doc program co-funded by the European Social Fund (ESF). The author benefited from the CERN Corresponding Associate Programme.



\begin{thebibliography}{99}

\bibitem{moedal-web} For general information on the MoEDAL experiment, see: \url{http://moedal.web.cern.ch/}

\bibitem{moedal-tdr} MoEDAL Collaboration, {\it Technical Design Report of the MoEDAL Experiment}, CERN Preprint, CERN-LHC-2009-006, MoEDAL-TDR-1.1 (2009), and references therein.

\bibitem{jim}  J.~L.~Pinfold,
  {\it J.\ Phys.\ Conf.\ Ser.\/}  {\bf 631} (2015) 012014.
  
\bibitem{LHC}  L.~Evans and P.~Bryant,
  {\it JINST} {\bf 3} (2008) S08001.

\bibitem{DeRoeck:2011aa}  A.~De Roeck, A.~Katre, P.~Mermod, D.~Milstead and T.~Sloan,
  {\it Eur.\ Phys.\ J.\/} C {\bf 72} (2012) 1985 [{\footnotesize\tt arXiv:1112.2999 [hep-ph]}].
  
\bibitem{Fairbairn07} M.~Fairbairn, A.~C.~Kraan, D.~A.~Milstead, T.~Sjostrand, P.~Z.~Skands and T.~Sloan,
  {\it Phys.\ Rept.\/} {\bf 438} (2007) 1 [{\footnotesize\tt hep-ph/0611040}]; \\
S.~Burdin, M.~Fairbairn, P.~Mermod, D.~Milstead, J.~Pinfold, T.~Sloan and W.~Taylor,
  {\it Phys.\ Rept.\/} {\bf 582} (2015) 1 [{\footnotesize\tt arXiv:1410.1374 [hep-ph]}].

\bibitem{Mitsou:2014dpa}  V.~A.~Mitsou [MoEDAL Collaboration],
  {\it EPJ Web Conf.\/}  {\bf 95} (2015) 04042 [{\footnotesize\tt arXiv:1411.7651 [hep-ph]}].
      
\bibitem{Acharya:2014nyr}    B.~Acharya {\it et al.}  [MoEDAL Collaboration],
  {\it Int.\ J.\ Mod.\ Phys.\/} A {\bf 29} (2014) 1430050 [{\footnotesize\tt arXiv:1405.7662 [hep-ph]}], and references therein.

\bibitem{LHCb-detector}   A.~A.~Alves, Jr. {\it et al.}  [LHCb Collaboration],
  {\it JINST} {\bf 3} (2008) S08005.
  
\bibitem{Joergensen:2012gy}    M.~D.~Joergensen, A.~De Roeck, H.-P.~Hachler, A.~Hirt, A.~Katre, P.~Mermod, D.~Milstead and T.~Sloan,
  {\it Searching for magnetic monopoles trapped in accelerator material at the Large Hadron Collider,}
  {\footnotesize\tt arXiv:1206.6793 [physics.ins-det]} (2012).
  
\bibitem{DeRoeck:2012wua}   A.~De~Roeck, H.~P.~Hachler, A.~M.~Hirt, M.-D.~Joergensen, A.~Katre, P.~Mermod, D.~Milstead and T.~Sloan,
  {\it Eur.\ Phys.\ J.\/} C {\bf 72} (2012) 2212.
 
\bibitem{mmt2012}   B.~Acharya {\it et al.} [MoEDAL Collaboration], 
  {\it Search for magnetic monopoles with the MoEDAL trapping detector in 8 TeV proton-proton collisions at the LHC,}
  to be submitted (2015).  
    
\bibitem{timepix}  N.~Asbah, C.~Leroy, S.~Pospisil and P.~Soueid,
  {\it JINST} {\bf 9} (2014) C05021.
  
\bibitem{atlas-monopole}   For magnetic monopoles, see e.g., G.~Aad {\it et al.} [ATLAS Collaboration],
  {\it Phys.\ Rev.\ Lett.\ } {\bf 109} (2012) 261803  [{\footnotesize\tt arXiv:1207.6411 [hep-ex]}]; 
  {\it Search for magnetic monopoles and stable particles with high electric charges in 8 TeV $pp$ collisions with the ATLAS detector,}
  {\footnotesize\tt arXiv:1509.08059 [hep-ex]} (2015).
  
\bibitem{Dirac1931kp}  P.~A.~M.~Dirac,
  {\it Proc.\ Roy.\ Soc.\ Lond.\/} A {\bf 133} (1931) 60.

\bibitem{Diracs_idea}   P.~A.~M.~Dirac,
  {\it Phys.\ Rev.\/} {\bf 74} (1948) 817.

\bibitem{tHooft-Polyakov} G.~'t Hooft,
  {\it Nucl.\ Phys.\/} B {\bf 79} (1974) 276; \\
A.~M.~Polyakov,
  {\it JETP Lett.\/} {\bf 20} (1974) 194
   [{\it Pisma Zh.\ Eksp.\ Teor.\ Fiz.\/} {\bf 20} (1974) 430].

\bibitem{Cho1996qd}  Y.~M.~Cho and D.~Maison,
  {\it Phys.\ Lett.\/} B {\bf 391} (1997) 360 [{\footnotesize\tt hep-th/9601028}]; \\ 
W.~S.~Bae and Y.~M.~Cho,
  {\it J.\ Korean Phys.\ Soc.\/} {\bf 46} (2005) 791 [{\footnotesize\tt hep-th/0210299}]; \\ 
Y.~M.~Cho, K.~Kim and J.~H.~Yoon, 
  {\it Mass of the Electroweak Monopole,} {\footnotesize\tt arXiv:1212.3885 [hep-ph]} (2012).

\bibitem{Vento2013jua}  V.~Vento and V.~S.~Mantovani,
  {\it On the magnetic monopole mass,}
  {\footnotesize\tt arXiv:1306.4213 [hep-ph]} (2013).
  
\bibitem{arttu} A.~Rajantie,
  {\it JHEP} {\bf 0601} (2006) 088  [{\footnotesize\tt hep-lat/0512006}].

\bibitem{khlopov} Y.~B.~Zeldovich and M.~Y.~Khlopov,
  {\it Phys.\ Lett.\/} B {\bf 79} (1978) 239.
  
\bibitem{Monopolium} C.~T.~Hill,
  {\it Nucl.\ Phys.\/} B {\bf 224} (1983) 469.

\bibitem{Monopolium1} V.~K.~Dubrovich,
  {\it Grav.\ Cosmol.\ Suppl.\/} {\bf 8N1} (2002) 122.

\bibitem{Rajantie:2012xh}   A.~Rajantie,
  {\it Contemp.\ Phys.\/} {\bf 53} (2012) 195  [{\footnotesize\tt arXiv:1204.3077 [hep-th]}].

\bibitem{laura}  L.~Patrizii and M.~Spurio,
  {\it Ann.\ Rev.\ Nucl.\ Part.\ Sci.\/} {\bf 65} (2015) 279  [{\footnotesize\tt arXiv:1510.07125 [hep-ex]}].

\bibitem{Epele0}  L.~N.~Epele, H.~Fanchiotti, C.~A.~Garcia Canal and V.~Vento,
  {\it Eur.\ Phys.\ J.\/} C {\bf 56} (2008) 87   [{\footnotesize\tt hep-ph/0701133}]; 
  {\it ibid.\/} {\bf 62} (2009) 587 [{\footnotesize\tt arXiv:0809.0272 [hep-ph]}].

\bibitem{Epele1} L.~N.~Epele, H.~Fanchiotti, C.~A.~G.~Canal, V.~A.~Mitsou and V.~Vento,
  {\it Eur.\ Phys.\ J.\ Plus} {\bf 127} (2012) 60    [{\footnotesize\tt arXiv:1205.6120 [hep-ph]}].

\bibitem{westmuckett} J.~R.~Ellis, N.~E.~Mavromatos and D.~V.~Nanopoulos,
  {\it Gen.\ Rel.\ Grav.\/}  {\bf 32} (2000) 943 [{\footnotesize\tt gr-qc/9810086}]; 
{\it Phys.\ Lett.\/} B {\bf 665} (2008) 412 [{\footnotesize\tt arXiv:0804.3566 [hep-th]}]; \\
J.~R.~Ellis, N.~E.~Mavromatos and M.~Westmuckett, 
 {\it Phys.\ Rev.\/} D \textbf{70} (2004) 044036 [{\footnotesize\tt gr-qc/0405066}];  
{\it ibid.\/} \textbf{71} (2005) 106006 [{\footnotesize\tt gr-qc/0501060}].
   
\bibitem{shiu}  G.~Shiu and L.~-T.~Wang,
  {\it Phys.\ Rev.\/} D {\bf 69} (2004) 126007 [{\footnotesize\tt hep-ph/0311228}].

\bibitem{Witten2002wb}  See, e.g., E.~Witten,
  {\it Comments on string theory,}
  {\footnotesize\tt hep-th/0212247} (2002), and references therein.  

\bibitem{Mavromatos:2010jt}   N.~E.~Mavromatos, S.~Sarkar and A.~Vergou,
  {\it Phys.\ Lett.\/} B {\bf 696}, 300 (2011) [{\footnotesize\tt arXiv:1009.2880 [hep-th]}]; \\
  N.~E.~Mavromatos, V.~A.~Mitsou, S.~Sarkar and A.~Vergou,
  {\it Eur.\ Phys.\ J.\/} C {\bf 72} (2012) 1956 [{\footnotesize\tt arXiv:1012.4094 [hep-ph]}].

\bibitem{mitsou-dm}  V.~A.~Mitsou,
  {\it Int.\ J.\ Mod.\ Phys.\/} A {\bf 28} (2013) 1330052  [{\footnotesize\tt arXiv:1310.1072 [hep-ex]}].

\bibitem{stauNLSP}  J.~R.~Ellis, K.~A.~Olive, Y.~Santoso and V.~C.~Spanos,
 {\it Phys.\ Lett.\/} B {\bf 565} (2003) 176  [{\footnotesize\tt arXiv:hep-ph/0303043}].

\bibitem{MC8} O.~Buchmueller {\it et al.},
  {\it Eur.\ Phys.\ J.\/} C {\bf 72} (2012) 2243 [{\footnotesize\tt arXiv:1207.7315 [hep-ph]}]; 
 {\it ibid.\/}  {\bf 74} (2014)  2922  [{\footnotesize\tt arXiv:1312.5250 [hep-ph]}].
    
\bibitem{Sato}  T.~Jittoh, J.~Sato, T.~Shimomura and M.~Yamanaka,
  {\it Phys.\ Rev.\/} D {\bf 73} (2006) 055009 [{\it Erratum-ibid.\/} D {\bf 87} (2013) 019901]] [{\footnotesize\tt hep-ph/0512197}].    
    
\bibitem{CAPTURE3} K.~Hamaguchi, M.~M.~Nojiri and A.~de Roeck,
  {\it JHEP} {\bf 0703} (2007) 046  [{\footnotesize\tt hep-ph/0612060}].
  
\bibitem{sleptonNLSP} J.~R.~Ellis, K.~A.~Olive and Y.~Santoso,
  {\it JHEP} {\bf 0810}  (2008) 005  [{\footnotesize\tt arXiv:0807.3736 [hep-ph]}].
  
\bibitem{stopNLSP}  J.~L.~Diaz-Cruz, J.~R.~Ellis, K.~A.~Olive and Y.~Santoso,
  {\it JHEP} {\bf 0705}  (2007) 003 [{\footnotesize\tt hep-ph/0701229}].
  
\bibitem{splitSUSY}  N.~Arkani-Hamed and S.~Dimopoulos, 
	{\it JHEP} {\bf 0506} (2005) 073 [{\footnotesize\tt hep-th/0405159}]; \\
G.~F.~Giudice and A.~Romanino,  
{\it Nucl.\ Phys.\/} B {\bf 699} (2004) 65 [{\it Erratum-ibid.\/} B {\bf 706} (2005) 65] [{\footnotesize\tt hep-ph/0406088}].
  
\end{thebibliography}
\end{document}